# Oxygen abundance in the Sloan Digital Sky Survey

F. Shi[1], X. Kong[1,2], F. Z. Cheng[1]

[1] Center for Astrophysics, University of Science and Technology of China, 230026, P. R. China
  e-mail: sfemail@mail.ustc.edu.cn
[2] National Astronomical Observatory, 2-21-1 Osawa, Mitaka, Tokyo 181-8588, Japan

**ABSTRACT**

*Context.*
*Aims.* To assess the possible systematic differences among different oxygen abundance indicators and understand the origin of nitrogen and the stars responsible for nitrogen production, we present two samples of H II galaxies from the Sloan Digital Sky Survey (SDSS) spectroscopic observations data release 3.
*Methods.* The electron temperatures($T_e$) of 225 galaxies are calculated with the photoionized H II model and $T_e$ of 3997 galaxies are calculated with an empirical method. The oxygen abundances from the $T_e$ methods of the two samples are determined reliably. The oxygen abundances from a strong line metallicity indicator, such as $R_{23}$, $P$, $N2$, and $O3N2$, are also calculated. We compared oxygen abundances of H II galaxies obtained with the $T_e$ method, $R_{23}$ method, $P$ method , $N2$ method, and $O3N2$method.
*Results.* The oxygen abundances derived with the $T_e$ method are systematically lower by ~0.2 dex than those derived with the $R_{23}$ method, consistent with previous studies based on H II region samples. No clear offset for oxygen abundance was found between $T_e$ metallicity and $P$, $N2$ and $O3N2$ metallicity. When we studied the relation between N/O and O/H, we found that in the metallicity regime of 12 + log(O/H) > 7.95, the large scatter of the relation can be explained by the contribution of small mass stars to the production of nitrogen. In the high metallicity regime, 12 + log(O/H) > 8.2, nitrogen is primarily a secondary element produced by stars of all masses.

## 1. Introduction

H II galaxies with strong emission lines are important probes for the formation and evolution of galaxies. Their spectra contain much important information needed to determine the star formation rate, initial mass function, element abundance, etc. (Stasinska & Leitherer 1996; Kennicutt 1998; Contini et al. 2002). The heavy element abundance is a key parameter for the formation and chemical evolution of a galaxy. Oxygen is one of the most important elements and is most easily and reliably determined since the most important ionization stages can all be observed. The oxygen abundance from the measurement of electron temperature from [O III]$\lambda\lambda$4959,5007/[O III]$\lambda$4363 is one of the most often-used methods. But [O III]$\lambda$4363 is usually weak and there are often large errors when measuring this line. Instead of the $T_e$ method, strong line methods such as the $R_{23}$ method, P method, $N2$ method or $O3N2$ method are used widely (Pagel et al. 1979; Kobulnicky, Kennicutt, & Pizagno 1999; Pilyugin et al. 2001; Charlot & Longhetti 2001; Denicoló et al. 2002; Pettini & Pagel 2003; Tremonti et al. 2004). Based on the homogeneous SDSS optical spectral sample, we determined the oxygen abundance of H II galaxies with both the $T_e$ method and these empirical methods. The results can be used to test the consistency of these different oxygen abundance indicators and to understand the physical origins of any systematic differences.

The N/O–O/H diagram is key for understanding the origin of nitrogen and the star formation history of galaxies. The origin of nitrogen has been discussed widely by many authors, but it is still uncertain in stars responsible for the production of nitrogen (Izotov & Thuan 1999; Kunth & Östlin 2000; Contini et al. 2002, Pilyugin et al. 2003). One limitation in studying the N/O–O/H relation is that it requires the measurement of [O III]$\lambda$4363 which is essential for the determination of the line temperature of nitrogen. To overcome this problem, Pilyugin et al. (2001) has presented a strong line method to calculate $T_e$ in place of [O III]$\lambda$4363 and confirmed this method in studying the origin of nitrogen (Pilyugin et al. 2003). We will use this method to study the relation of the N/O–O/H in a large H II galaxy sample from SDSS DR3.

The Sloan Digital Sky Survey (SDSS) is the most ambitious imaging and spectroscopic survey to date, and will eventually cover a quarter of the sky (York et al. 2000). The large area coverage and moderately deep survey limit of the SDSS make it suitable to study the large-scale structure and the characteristics of the galaxy population in the local universe. Because of its homogeneity, the SDSS provides a large sample of H II galaxies where oxygen abundance and N/O ratio can be calculated with the classic $T_e$ method.

Based on the SDSS DR3 starburst spectral sample, we present a sample whose $T_e$ can be determined reliably (Sect. 2). The oxygen abundance of the sample is calculated with the $T_e$ method and strong line methods (Sect. 3). The results can be used to test the consistency of these different oxygen abundance indicators and to understand the physical origins of any systematic differences(section 4). The nitrogen abundance are calculated and the origin of nitrogen is discussed in Sect. 4. The conclusions are summarized in Sect. 5.

## 2. Data sample

### 2.1. SDSS DR3

The SDSS targets two samples of galaxies: a flux-limited sample to $r = 17.77$, called the main galaxy sample (Strauss et al. 2002), and a flux- and color-selected sample extending to $r = 19.5$, designed to target luminous red galaxies (LRGs, Eisenstein et al. 2001) assigned to spectroscopic plates by an adaptive tiling algorithm (Blanton et al. 2003).The galaxies are observed with a pair of fiber-fed CCD spectrographs, after which the spectroscopic data reduction and redshift determination are performed by automated pipelines. The spectra are in the wavelength range from 3800 Å to 9200 Å with spectral resolution $R \sim 1800$. The rms galaxy redshift errors are ∼30 km$^{-1}$. The fibers have a diameter of 3″.

In this paper we used the reduced spectra from the DR3 database (Abazajian et al. 2005), containing ∼ 370,000 galaxy spectra and covering a total sky area of ∼4188 deg$^2$. These spectra have been downloaded from the official DR3 web-site (http://www.sdss.org/dr3/), and are used as the primary database in our work.

### 2.2. Data reduction

Subtracting the underlying stellar absorption is crucial to study nebular emission lines and the abundances of gaseous material (Kong et al. 2003). In this paper we use the method developed by Li et al. (2005) to subtract the underlying starlight. Briefly, a set of absorption-line templates with zero velocity dispersion is constructed, on the basis of two successive applications of the technique of Principal Component Analysis (PCA), first to 204 stars in the stellar library STELIB (Le Borgne et al. 2003), then to a uniform sample of galaxies selected from SDSS DR1. The spectra of all galaxies in SDSS DR3 are then fitted with the templates with iterative rejection of emission lines and bad pixels. For each spectrum, the best-fitting model is subtracted from the original spectrum, yielding a pure emission-line spectrum from which the emission-line parameters (the central wavelength, line intensity and line width) are measured. Extensive tests have shown that the spectra of different types of galaxies can be modeled quite accurately using this method (see Li et al. 2005 for more detailed discussion).

The extinction of interstellar dust in star-forming galaxies modifies the spectra of these objects. It is necessary to correct all observed line fluxes for this internal reddening. The most widely used method to correct the emission line spectra for the presence of dust is based on the relative strengths of low order Balmer lines. In order to have an internally consistent sample, we applied this method to each of our galaxies, using the ratio of the two strongest Balmer lines, H$\alpha$/H$\beta$ ( We assume the theoretical H$\alpha$/H$\beta$ = 2.86 at 10,000 K). Galactic extinction and underlying stellar absorption were corrected and the effective absorption curve $\tau_\lambda = \tau_V(\lambda/5500\text{Å})^{-0.7}$, which was introduced by Charlot & Fall (2000), was used to make this correction.

### 2.3. Sample selection

Within our primary starburst sample, we made use of the spectral diagnostic diagrams from Kauffmann et al. (2003) to classify galaxies as either star-forming galaxies (SFGs), active galactic nuclei (AGN), or unclassified. The diagram used was [O III]$\lambda$5007/H$\beta$ vs [N II]$\lambda\lambda$6548, 6583/H$\alpha$.

$$\log([\text{OIII}]/\text{H}\beta) > 0.61/(\log([\text{NII}]/\text{H}\alpha) - 0.05) + 1.3 \qquad (1)$$

To determine the $T_e$ metallicity reliably, we selected our H II galaxy sample with flux uncertainties for [O III]$\lambda$4363, [O III]$\lambda$4959, [O III]$\lambda$5007, [O II]$\lambda$3727, [N II]$\lambda$6583, H$\alpha\lambda$6563, H$\beta\lambda$4861 larger than 5$\sigma$. Because the [O III]$\lambda$4363 line is only strong in metal-poor galaxies(12 + log(O/H) < 8.2), only galaxies that satisfy $log_{10}([\text{N II}]\lambda6583]/\text{H}\alpha) < -1.26$ are included in our sample. We use 225 metal poor galaxies in our first sample (**Sample I**).

Our first sample is rather small compared to the SDSS data. One reason is that the temperature sensitive line [O III]$\lambda$4363 is only strong in metal-poor galaxies (12 + log(O/H) < 8.2) and the number of metal-poor galaxies in the universe is relatively small compared to the metal-rich galaxies. To obtain as large a sample as possible in which $T_e$ metallicity can be determined, an empirical relationship between $T_e$ and the strong line ratio (Pilyugin 2001) in the high metallicity region (12 + log(O/H) > 8.2) can be used to replace the weak [O III]$\lambda$4363 line to calculate the $T_e$ of H II galaxies. For this method, a second sample is derived based on the prerequisite that the flux of [O III]$\lambda$4959, [O III]$\lambda$5007, [O II]$\lambda$3727, [N II]$\lambda$6583, H$\alpha\lambda$6563 and H$\beta\lambda$4861 lines is greater than 5 times the flux uncertainty. We used 3997 starburst galaxies as **Sample II**.

## 3. Determination of oxygen abundance

### 3.1. $T_e$ method

To derive oxygen abundances with the $T_e$ method, we determine $T_e$ and $n_e$ for a two-zone photoionized H II region model. It is well established that for low metallicity galaxies, 12 + log(O/H) < 8.2, [O III]$\lambda$4363 is prominent and can be measured accurately, while for high metallicity galaxies, [O III]$\lambda$4363 is weak and the error of its measurement is large. We use an $N2$ indicator to distinguish high metallicity regions from low metallicity regions (see Sect. 3.4). For low metallicity galaxies (12 + log(O/H)$_{N2}$ < 8.2), we use a five-level statistical equilibrium model in the IRAF NEBULAR package (de Robertis, Dufour, & Hunt 1987; Shaw & Dufour 1995), which makes use of the latest collision strengths and radiative transition probabilities to determine the $T_e$ and $n_e$. For high metallicity galaxies (12 + log(O/H)$_{N2}$ > 8.2), an empirical relation of $T_e$ and strong spectral lines has been adopted for the electron temperature determination (Pilyugin 2001). The temperature will be used to derive of the O$^{+2}$ ionic abundances.

To estimate the temperature in the low-temperature zone $T_e$(O II), the relation between $T_e$(O II) and $T_e$(O III) from Garnett (1992) is utilized:

$$t_e(\text{O II}) = 0.7 \times t_e(\text{O III}) + 0.3, \quad (2)$$

where $t_e = T_e/10^4$ K. The temperature $T_e$(O II) is used to derive the O$^+$ ionic abundance.

After calculation of $T_e$ and $n_e$ for the high-temperature zone and low-temperature zone, we use the expressions from Pagel et al. (1992) to calculate the oxygen abundance for these two zones. Then we simply sum O$^+$ and O$^{++}$ as our final oxygen abundance.

$$\frac{\text{O}}{\text{H}} = \frac{O^+}{H^+} + \frac{O^{++}}{H^+}, \quad (3)$$

$$12 + \log(O^{++}/H^+) = \log \frac{I([\text{O III}]\lambda\lambda 4959, 5007)}{I(\text{H}\beta)} +$$
$$6.174 + \frac{1.251}{t_e(\text{O III})} - 0.55 \log t_e(\text{O III}), \quad (4)$$

$$12 + \log(O^+/H^+) = \log \frac{I([\text{O II}]\lambda 3727)}{I(\text{H}\beta)} + 5.890 +$$
$$\frac{1.676}{t_e(\text{O II})} - 0.40 \log t_e(\text{O II}) + \log(1 + 1.35x), \quad (5)$$

where $n_e$ is the electron density in cm$^{-3}$, and $x = 10^{-4} n_e t_e(\text{O II})^{-1/2}$.

### 3.2. $R_{23}$ method

$T_e$ metallicity determination requires the accurate measurement of the weak auroral forbidden emission line [O III]$\lambda$4363. The flux intensity of [O III]$\lambda$4363 strongly anticorrelates with the abundance of galaxies. Its flux intensity becomes undetectable in high metallicity galaxies (12 + log(O/H) > 8.2).

For this reason, the strong line metallicity indicator $R_{23}$ has been developed since Pagel et al. (1979) introduced it for the first time. We use the most recent $R_{23}$ analytical calibrations given by Kobulnicky et al. (1999) which are based on the models by McGaugh (1991) to determine the oxygen abundances in our sample.

The major difficulty associated with this method is that the relation between oxygen abundance and $R_{23}$ is double valued, requiring some assumption or rough a priori knowledge of a galaxy's metallicity in order to locate it on the appropriate branch of the curve. In this work, the [N II]$\lambda$6583/H$\alpha$ line ratio will be used to break the degeneracy of the $R_{23}$ relation (Denicoló et al. 2002). The division between the upper and the lower branch of the $R_{23}$ relation occurs around log([N II]$\lambda$6583/H$\alpha$) ≃ −1.26 (12 + log(O/H) ≃ 8.2).

### 3.3. P method

The $R_{23}$ method was used widely but $R_{23}$ abundances were found to be systematically larger than the $T_e$ metallicity. Pilyugin (2000, 2001) found that its error was two parts, a random error and a systematic error. The origin of this systematic error is the dependence of the oxygen emission lines on not only the oxygen abundance, but also on the other physical conditions (hardness of the ionizing radiation and a geometric factor). Pilyugin (2000, 2001) introduced the $P$ method, another strong line metallicity indicator to overcome these problems. We use the expression of Pilyugin (2001) to calculate the abundance of oxygen in high metallicity regions (log([N II]$\lambda$6583/H$\alpha$) > −1.26, or 12 + log(O/H) > 8.2):

$$12 + \log(\text{O/H})_P = \frac{R_{23} + 54.2 + 59.45P + 7.31P^2}{6.07 + 6.71P + 0.37P^2 + 0.243R_{23}} \quad (6)$$

and in low metallicity regions (log([N II]$\lambda$6583/H$\alpha$) < −1.60, or 12 + log(O/H) < 7.95):

$$12 + \log(\text{O/H}) = 6.35 + 1.45 \log R_3 - 3.19 \log P, \quad (7)$$

where $R_3 = I([\text{O III}]\lambda\lambda 4959, 5007)/I(\text{H}\beta)$, and $P = R_3/R_{23}$.

### 3.4. $N2$ method

Both $R_{23}$ and $P$ metallicity are double valued. It is instructive to use one metallicity to describe the whole metallicity with a single slope. The $N2 \equiv \log[I([\text{N II}]\lambda 6583)/I(\text{H}\alpha)]$ index was found to fulfill this requirement by Denicoló et al. (2002). A least squares fit to the data simultaneously minimizing the errors in both axes gives

$$12 + \log(\text{O/H}) = 9.12 + 0.73 \times N2. \quad (8)$$

The $N2$ indicator has advantages superior to the other metallicity indicator. The $N2$ vs. metallicity relation is monotonic, and the $N2$ line ratio does not depend on reddening corrections or flux calibration. These advantages make $N2$ indicators able to break the degeneracy of the $R_{23}$–(O/H) (in Sect. 3.2) and the $P$-(O/H) (in Sect. 3.3) relation.

### 3.5. $O3N2$ method

$O3N2 \equiv \log \{[I([\text{O III}]\lambda 5007)/I(\text{H}\beta)] / [I([\text{N II}]\lambda 6583)/I(\text{H}\alpha)]\}$ is another indicator that is monotonic. It was introduced by Alloin et al. (1979) and further studied by Pettini & Pagel (2004). Pettini & Pagel (2004) found that at $O3N2 \leq 1.9$, there appears to be a relatively tight, linear and steep relationship between $O3N2$ and log(O/H). A least squares linear fit to the data in the range −1 < $O3N2$ < 1.9 yields the relation:

$$12 + \log(\text{O/H}) = 8.73 - 0.32 \times O3N2. \quad (9)$$

We use this expression to calculate the $O3N2$ oxygen abundance.

## 4. Results

### 4.1. Comparison with previous studies

There exist previous SDSS results that studied the abundance of starburst galaxies using the $T_e$ method. Kniazev et al. (2004) presented the first edition of the SDSS H II galaxies with oxygen abundances obtained with the classic $T_e$ method (7.6 < 12 + log(O/H) < 8.4). Tremonti et al. (2004) estimated metallicity statistically for ∼ 45,000 star-forming galaxies, based on simultaneous fits of all the most prominent emission lines ([O II], H$\alpha$, [O III], H$\beta$, [N II], [S II]) with a model designed for the interpretation of integrated galaxy spectra (Charlot & Longhetti 2001). The catalogs of Kniazev et al. (2004) and Tremonti et al. (2004) coincide partly with ours. Because the $T_e$ method is the most accurate and reliable way to calculate oxygen abundance (better than 0.1 dex), it is instructive to compare our metallicity with Kniazev et al. (2004).

Figure 1 shows the comparison of our oxygen abundance with Kniazev et al. (2004) and Tremonti et al. (2004). It is clear that our abundances are systematically larger than Kniazev's results. In high metallicity regions (12 + log(O/H) > 8.2), the difference between them is especially large (0.1–0.4 dex). The difference may be induced by the inaccuracy of measurements for the weak [O III]$\lambda$4363 line in high metallicity regions. No galaxy in the moderate metallicity regime (8.2 > 12 + log(O/H) > 7.95) is found at the same time both in Kniazev's sample and ours. In the Kniazev sample, $T_e$ in high metallicity regions are derived from the noisy [O III]$\lambda$4363 line. In our sample, $T_e$ in high metallicty regions (12 + log(O/H) > 8.2) are derived with the P method (Pilyugin 2001) and in low metallicity regions (12 + log(O/H) < 8.2), $T_e$ are derived from [O III]$\lambda$4363. Melbourne et al. (2004) argued that the $T_e$ method using $T_e$ from [O III]$\lambda$4363 can underestimate the metallicity of a galaxy when [O III]$\lambda$4363 is noisy. In this circumstance, $T_e$ metallicity using $T_e$ from the P method is a better choice.

The difference between our $T_e$, $R_{23}$ metallicity and the metallicity of Tremonti et al. (2004) is large in the whole metallicity regime. Further research on H II models in high metallicity galaxies is needed to solve this disagreement.

### 4.2. Oxygen abundance from different indicators

Both the $T_e$ method and other strong line methods are used widely to derive the abundance of H II galaxies. The $T_e$ method is the most reliable method but it suffers the problem of weakness of [O III]$\lambda$4363. The $R_{23}$ method and $P$ method is double valued. $N2$ and $O3N2$ metallicity has a large dispersion. In consideration of this, it is instructive to study the difference of the oxygen abundance derived with the $T_e$-based method and the empirical method.

In Figure 2, we shows the plot of the $T_e$-based oxygen abundance against the empirical method oxygen abundances. Galaxies with temperatures determined from the temperature sensitive line [O III]$\lambda$4363 are presented as red crosses, and those with temperatures from the strong spectral lines (following Pilyugin et al. 2001, see Sect. 3 of this paper) are presented as black filled circles. Both abundances with temperatures determined from the [O III]$\lambda$4363 and from the strong spectral lines are regarded as the standard $T_e$–based abundance and used to study the relation between the P–, $R_{23}$–, $N2$– and $O3N2$– method abundances and the $T_e$–based abundances.

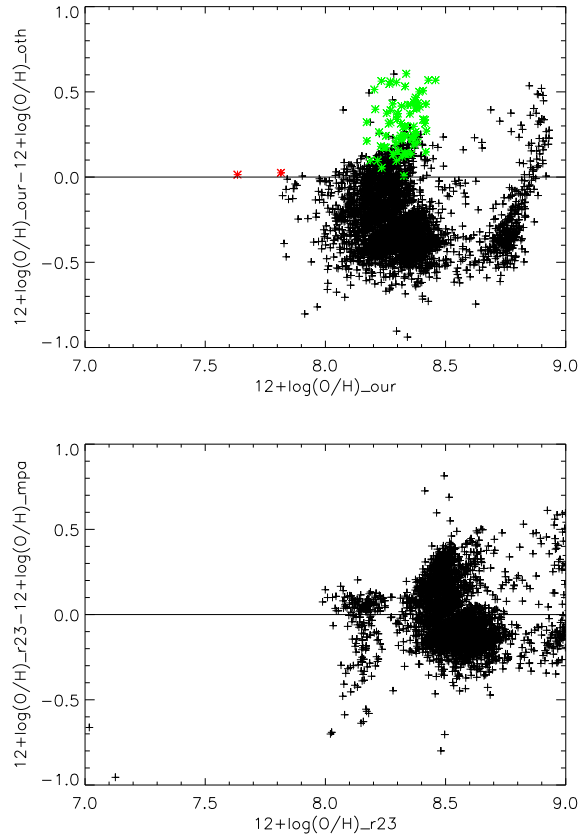

**Fig. 1.** Upper panel: Comparison of the electron temperature-based oxygen abundances with those derived by Kniazev et al. (2004) (green stars are the galaxies for which our $T_e$ are derived from the P method and the red stars are the galaxies for which our $T_e$ are derived from [O III]$\lambda$4363) and the median of the likelihood distribution for 12 + log(O/H) in Tremonti et al. (2004) (black crosses) is given. Lower panel: Comparison of the oxygen abundances from the $R_{23}$ method with the median of the likelihood distribution for 12 + log(O/H) in Tremonti et al. (2004) (black crosses).

Some $T_e$ metallicities from the P method are in the range of 12 + log(O/H) < 8.2, although this range is not defined for the P method. It is caused by the scatter of the $N2$ vs. O/H relation such that objects close to $N2 = −1.26$ may be either in the high or low metallicity branch of the $P$ calibrations. When analyzing the galaxies in our sample, we found that most of them have high or low metallicity, with only ∼ 1 % of them close to $N2 = −1.26$. Therefore, the scatter of the $N2$ vs. O/H relation does not affect the $T_e$ abundances from the $P$ method for most galaxies in our sample.

Although having a large scatter, there is an agreement between the oxygen abundance from the $T_e$ method and those from the empirical methods. There is a residual of $\sim 0.2$ dex between $R_{23}$ metallicity and $T_e$ metallicity, while there is no clear residual between P, N2, O3N2 metallicity and $T_e$ metallicity for either Sample I and Sample II.

If the $T_e$ abundances are correct, it implies that most studies of the $R_{23}$ galactic abundances have over-estimated the true absolute oxygen abundances by factors of $\sim 1.8$ for high metallicity galaxies, as was noted by Kennicutt et al. (2003). There is a good agreement between the O/H$_{T_e}$ and the O/H$_P$ abundances for high metallicity galaxies. Because the temperature of these high metallicity galaxies was determined by the $P$ method too (Pilyugin 2001), this agreement can be understood easily, and cannot be used to test the validity of the $P$ method. The relation between $P$ metallicity and $T_e$ metallicity derived here is similar to the result of Shi et al. (2005) from the study of oxygen abundance for 72 blue compact galaxies. For N2 and O3N2 metallicity, Shi et al. (2005) found that there is also a residual between N2, O3N2 metallicity and $T_e$ metallicity(0.20 dex and 0.09 dex respectly) with less dispersion than the SDSS sample here. Because of the large dispersion for Sample I and II, we cannot determine if there is no systematic residual between the $T_e$ method and the $N2$, $O3N2$ metallicity or if the residuals are hidden by the scatter.

Many galaxies show a strong deviation of $\sim 0.4$ dex between $T_e$ metallicity and $N2$, $O3N2$ metallicity. The reason is believed to be related to the evolutionary state of the starbursts. Such galaxies seem to be past the peak of star formation and so have an elevated [O III]$\lambda 3727$/H$\beta$ vs [N II]$\lambda 6583$/H$\alpha$. The enhanced [N II]$\lambda 6583$/H$\alpha$ make the $N2$ metallicity higher. The increment of [O III]$\lambda 5007$/H$\beta$ makes the $O3N2$ metallicity higher.

For $R_{23}$, $P$, $N2$ and $O3N2$ metallicity, the residual between the strong line metallicity of Sample II and $T_e$ metallicity seems to be greater than that of Sample I. The reason may reside in the calibration of $T_e$ by the P method. Other discussions of the nature of the relation between the $T_e$ method and other strong line methods can be found in Kennicutt, Bresolin & Garnett (2003).

### 4.3. The relation between N/O and O/H

The origin of nitrogen and the stars responsible for nitrogen production has been discussed widely by many authors (Izotov et al. 1999; Kunth & Östlin 2000; Pilyugin et al. 2003). Their conclusions can be summarized as follows. For galaxies with $12 + \log(O/H) < 7.6$, primary nitrogen is believed to be produced by massive stars only. In this scenario, galaxies in that O/H range are young, so that primary nitrogen from intermediate mass stars has not been released yet. Only in slightly older galaxies, do intermediate-mass stars start to contribute secondary nitrogen, leading to the higher mean and scatter of N/O ratios seen for $7.6 < 12 + \log(O/H) < 8.2$. For an alternative interpretation, see e.g. Izotov et al. (2004) and references therein. For the galaxies with high metallicity, $12 + \log(O/H) > 8.2$, the N/O ratio increases with the oxygen abundance more rapidly, indicating that, in this metallicity regime, nitrogen is primarily a *secondary* element, and the contribution from the primary production is not significant. In the case of *secondary* synthesis, oxygen and carbon have been produced in the previous generations of stars, and the nitrogen, produced in the present generation of stars, should be proportional to their initial heavy element abundance. Secondary nitrogen production is expected in stars of all masses (see Izotov & Thuan 1999). In the case of *primary* nitrogen synthesis, on the other hand, oxygen and carbon are produced in the same stars prior to the CNO cycle rather than in previous generations, and nitrogen production should be independent of the initial heavy element abundance. Primary nitrogen production is thought to occur mainly in intermediate-mass stars, yet important contributions may also come from high mass stars (see for example Weaver & Woosley 1995, Izotov & Thuan 1999, Izotov et al. 2004, and references therein).

To understand the origin of nitrogen in our sample, we plot the distributions of $12 + \log(O/H)$ and $\log(N/O)$ abundance ratios for our sample galaxies in Figure 3. The N/O abundance ratios in SFGs were determined from the expression (Pagel et al. 1992):

$$\log(N/O) = \log(N^+/O^+) = \log\frac{I([\text{N{\sc ii}}]\lambda\lambda 6548, 6584)}{I([\text{O{\sc ii}}]\lambda 3727)}$$
$$+0.31 - \frac{0.726}{t_e([\text{N{\sc ii}}])} - 0.02\log t_e([\text{N{\sc ii}}]) - \log\frac{1+1.35x}{1+0.12x}, \quad (10)$$

where $t_e([\text{N{\sc ii}}]) = t_e([\text{O{\sc ii}}])$, and $x = 10^{-4}n_e t_e([\text{N{\sc ii}}])^{-1/2}$.

Figure 3 give the expected result, as discussed above, that the ratio of nitrogen to oxygen increases in moderate metallicity regions ($7.6 < 12 + \log(O/H) < 8.2$) and increases more rapidly in the high metallicity region ($12 + \log(O/H) > 8.2$). This result may imply that the primary plus secondary origin from massive and intermediate-mass stars in the metallicity range of $7.6 < 12 + \log(O/H) < 8.2$ and secondary origin from stars of all masses dominates in the metallicity range of $12 + \log(O/H) > 8.2$. The scatter of the N/O–O/H relation is quite large, which can be explained by a time delay of nitrogen production, selective heavy-element loss through enriched galactic winds and different star formation histories in galaxies(Izotov & Thuan 1999; Pilyugin et al. 2003).

Kauffmann et al.(2003) shows that galaxies with the concentration value $C > 2.6$ are mostly early type galaxies whereas late type galaxies have $C < 2.6$. It is well known that early type galaxies are dominated by old/small mass stars and late type galaxies are dominated by young/massive stars. Pilyugin et al. (2003) showed from their study that small-mass stars can contribute to the production of nitrogen with long time delays (several Gyr). To examine this view in our sample, we replot the mean relation of N/O–O/H in the lower panel of Figure 3. The black, red and green lines indicate the mean N/O ratio at a given O/H for the subsample of $C > 2.6$, $2.3 < C < 2.6$, $2.0 < C < 2.3$. Figure 3 shows that the early type galaxies have larger N/O than late type glaxies at the given O/H in the metallicity regime of $12 + \log(O/H) > 7.95$. Our result supports the idea that the high nitrogen values in the early type galaxies can be caused by the contribution of small-mass stars to the production of nitrogen in the metallicity regime of $12 + \log(O/H) > 7.95$, which agrees with Pilyugin et al. (2003).

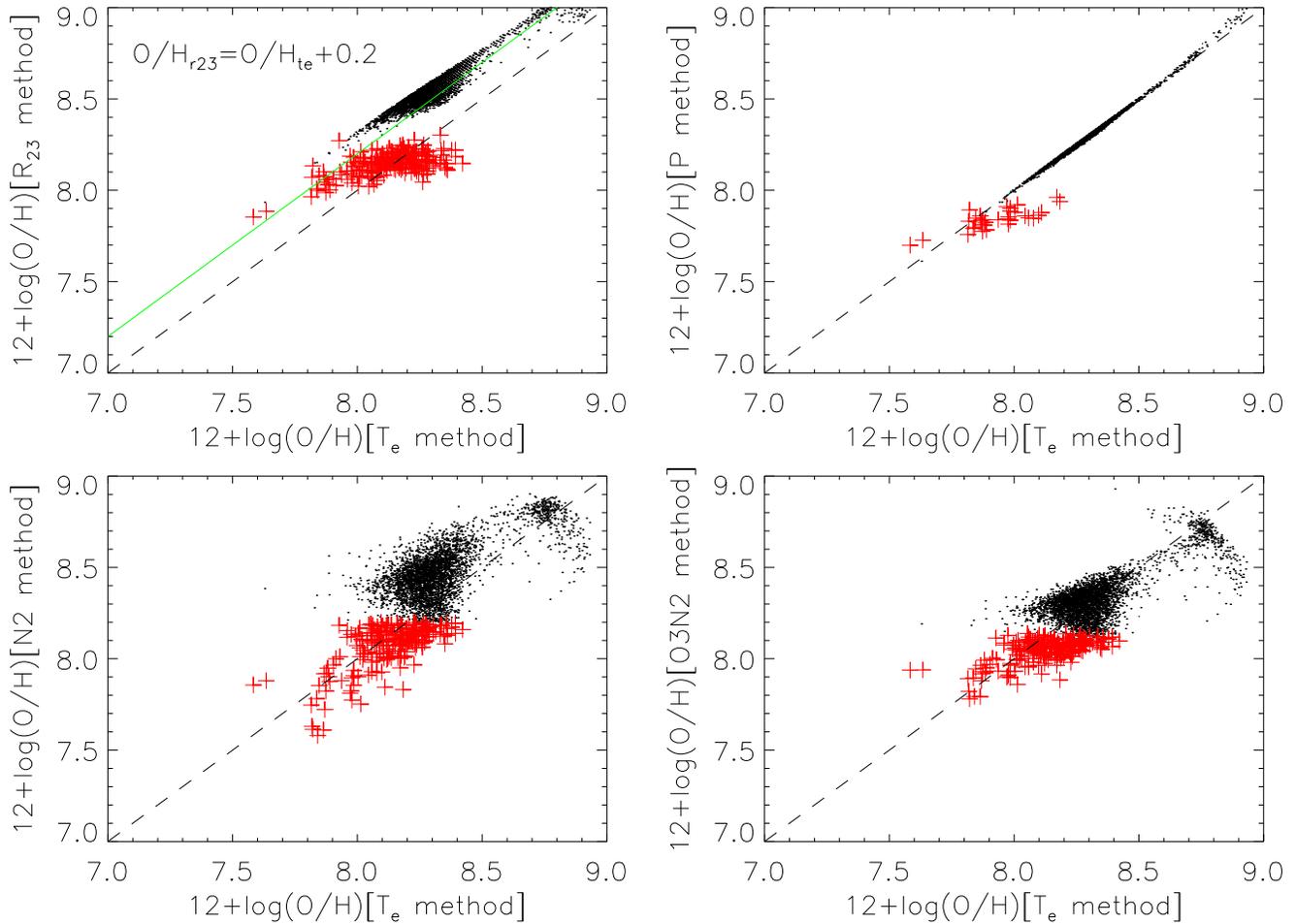

**Fig. 2.** Comparison of the electron temperature-based abundances with those derived from different strong line indicators. Sample I red crosses, and Sample II black filled circles.

## 5. Conclusions

We present a large sample from the H II galaxies in SDSS DR3 and have determined the electron temperatures, electron densities, nitrogen abundance and oxygen abundance for the sample, using different oxygen abundance indicators. We obtained the following results.

1. The oxygen abundances in our sample of HII galaxies range from 7.58 to 8.93. The oxygen abundance derived from the $T_e$ based method is in good agreement with the values of previous studies using the $T_e$ method, especially in the low metallicity regime. The oxygen abundance derived from the $T_e$ based method and strong line ratio methods differ considerably from the values derived from H II models. More detailed investigation of H II models is needed to resolve these inconsistencies.
2. The relations between the oxygen abundances from different indicators were investigated and the expected results are found. The $R_{23}$ method is systematically 0.2 dex higher than the $T_e$ method. $P$ method values agree with the $T_e$ method within a small error. $N2$ and $O3N2$ methods are in agreement with the $T_e$ method although there is large scatter between them.
3. The relation between the N/O and the O/H from our sample confirms the results of Izotov & Thuan (1999) and Pilyugin et al. (2003). In the metallicity regime $12 + \log(O/H) > 7.95$, the large scatter of the N/O–O/H relation can be explained by the contribution of small mass stars to the production of nitrogen. In the high metallicity regime, $12 + \log(O/H) > 8.2$, nitrogen is primarily a secondary element produced by stars of all masses.

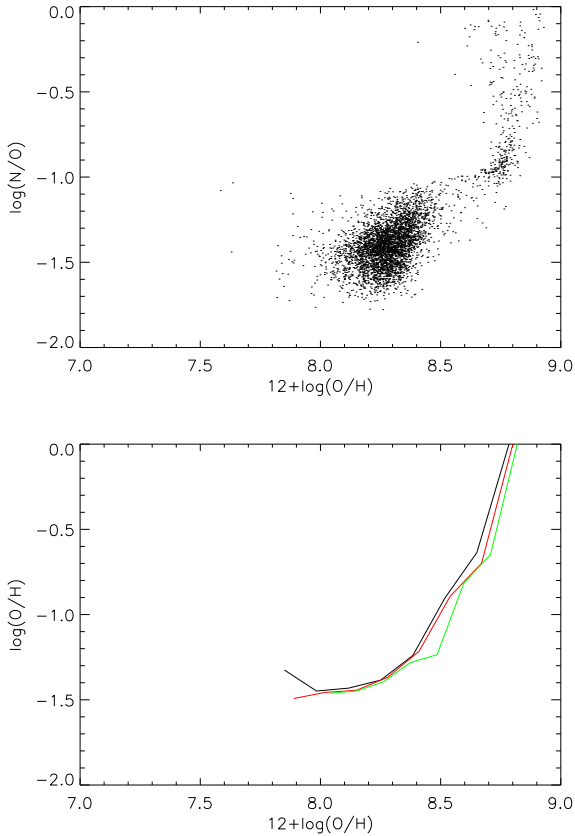

**Fig. 3.** Upper panel: Logarithm of the number ratio of nitrogen to oxygen abundances plotted against 12 + log(O/H). Lower panel: Logarithm of the mean value of the ratio of nitrogen to oxygen abundances plotted against 12 + log(O/H). The black, red and green lines indicate the subsamples of $C > 2.6$, $2.3 < C < 2.6$, $2.0 < C < 2.3$.